\documentclass[aps,twocolumn,superscriptaddress,amsmath,amssymb]{revtex4-2}
\usepackage{graphicx}
\usepackage{dcolumn}
\usepackage{bm}
\usepackage{braket}
\usepackage{appendix}
\usepackage{subfigure}
\usepackage{hyperref}
\usepackage{color}
\usepackage{xcolor}
\usepackage{amsmath}
\usepackage{nccmath}
\usepackage[english]{babel}
\usepackage{svg}
\usepackage{graphicx,float}

\begin{document}

\title[Nuclear Spin in hBN]
 {Nuclear spin polarization and control in a van der Waals material}
 
\author{Xingyu Gao}
\affiliation{Department of Physics and Astronomy, Purdue University, West Lafayette, Indiana 47907, USA}
\author{Sumukh Vaidya}
\affiliation{Department of Physics and Astronomy, Purdue University, West Lafayette, Indiana 47907, USA}
\author{Kejun Li}
\affiliation{Department of Physics, University of California, Santa Cruz, CA, 95064, USA}
\author{Peng Ju}
\affiliation{Department of Physics and Astronomy, Purdue University, West Lafayette, Indiana 47907, USA}
\author{Boyang Jiang}
\affiliation{Elmore Family School of Electrical and Computer Engineering, Purdue University, West Lafayette, Indiana 47907, USA}
\author{Zhujing Xu}
\affiliation{Department of Physics and Astronomy, Purdue University, West Lafayette, Indiana 47907, USA}
\author{Andres E. Llacsahuanga Allcca}
\affiliation{Department of Physics and Astronomy, Purdue University, West Lafayette, Indiana 47907, USA}
\author{Kunhong Shen}
\affiliation{Department of Physics and Astronomy, Purdue University, West Lafayette, Indiana 47907, USA}
\author{Takashi Taniguchi}
\affiliation{International Center for Materials Nanoarchitectonics, 
National Institute for Materials Science,  1-1 Namiki, Tsukuba 305-0044, Japan}
\author{Kenji Watanabe}
\affiliation{Research Center for Functional Materials, 
National Institute for Materials Science, 1-1 Namiki, Tsukuba 305-0044, Japan}
\author{Sunil A. Bhave}
\affiliation{Elmore Family School of Electrical and Computer Engineering, Purdue University, West Lafayette, Indiana 47907, USA}
\affiliation
{Purdue Quantum Science and Engineering Institute, Purdue University, West Lafayette, Indiana 47907, USA}
\affiliation
{Birck Nanotechnology Center, Purdue University, West Lafayette,
	IN 47907, USA}
\author{Yong P. Chen}
\affiliation{Department of Physics and Astronomy, Purdue University, West Lafayette, Indiana 47907, USA}
\affiliation{Elmore Family School of Electrical and Computer Engineering, Purdue University, West Lafayette, Indiana 47907, USA}
\affiliation
{Purdue Quantum Science and Engineering Institute, Purdue University, West Lafayette, Indiana 47907, USA}	
\affiliation
{Birck Nanotechnology Center, Purdue University, West Lafayette,
	IN 47907, USA}
\affiliation{WPI-AIMR International Research Center for Materials Sciences, Tohoku University, Sendai 980-8577, Japan}
\author{Yuan Ping}
\affiliation{Department of Chemistry and Biochemistry, University of California, Santa Cruz, CA, 95064, USA}
\author{Tongcang Li}
\email{tcli@purdue.edu}
\affiliation{Department of Physics and Astronomy, Purdue University, West Lafayette, Indiana 47907, USA}
\affiliation{Elmore Family School of Electrical and Computer Engineering, Purdue University, West Lafayette, Indiana 47907, USA}
\affiliation
{Purdue Quantum Science and Engineering Institute, Purdue University, West Lafayette, Indiana 47907, USA}
\affiliation
{Birck Nanotechnology Center, Purdue University, West Lafayette,
	IN 47907, USA}

\date{\today}



\begin{abstract}
{\normalsize Van der Waals layered materials are a focus of materials research as they support strong quantum effects and can easily form heterostructures. Electron spins in van der Waals materials played crucial roles in many recent breakthroughs, including topological insulators, two-dimensional (2D) magnets, and spin liquids. However, nuclear spins in van der Waals materials remain an unexplored quantum resource. Here we report the first demonstration of optical polarization and coherent control of nuclear spins in a van der Waals material at room temperature. We use negatively-charged boron vacancy ($V_B^-$) spin defects in hexagonal boron nitride to polarize nearby nitrogen nuclear spins.  Remarkably, we observe the Rabi frequency of nuclear spins at the excited-state level anti-crossing of  $V_B^-$ defects to be 350 times larger than that of an isolated nucleus, and demonstrate  fast coherent control of nuclear spins. 
We also detect strong electron-mediated nuclear-nuclear spin coupling that is 5 orders of magnitude larger than the direct nuclear spin dipolar coupling, enabling multi-qubit operations. Nitrogen nuclear spins in a triangle lattice will be suitable for large-scale quantum simulation. Our work opens a new frontier with nuclear spins in van der Waals materials for quantum information science and technology. 
}
\end{abstract}

\maketitle


Since the discovery of graphene, van der Waals (vdW) layered materials have been a focus of materials research for the last two decades \cite{doi:10.1126/science.1102896,zhang2005experimental,cao2018unconventional,geim2013van,novoselov20162d}. Thanks to their weak inter-layer interaction, vdW materials can be readily exfoliated and integrated with different materials and structures \cite{geim2013van,novoselov20162d}. Electron spins in vdW materials played essential roles in recent development in spintronics and condensed-matter physics, including  topological insulators \cite{PhysRevLett.95.146802,PhysRevLett.95.226801,RevModPhys.82.3045}, two-dimensional (2D) magnets \cite{gong2017discovery,huang2017layer}, and spin liquids \cite{banerjee2016proximate,chen2020strong}. Most vdW materials also have non-zero nuclear spins, which will have applications in  quantum sensing and quantum information processing if they can be polarized efficiently and controlled coherently \cite{liu20192d,kane1998silicon,cai2013large}. Nuclear spins in liquids have been used to perform quantum algorithms with conventional nuclear magnetic resonance (NMR) systems \cite{gershenfeld1997bulk}. However,  the thermal polarization of nuclear spins is extremely low in realistic magnetic fields at room temperature because of their small gyromagnetic ratio \cite{gershenfeld1997bulk}. Recently, it was theoretically proposed to couple a 2D lattice of nuclear spins to a diamond nitrogen-vacancy (NV) center for large-scale quantum simulation \cite{cai2013large}. But the achieved coupling has been too weak to use a diamond NV center to polarize nuclear spins in a vdW material so far \cite{lovchinsky2017magnetic}.  To our best knowledge, there is still no report on efficient polarization and coherent control of nuclear spins in a vdW material.

\begin{figure*}[htbp]
	\centering
	\includegraphics[width=1\textwidth]{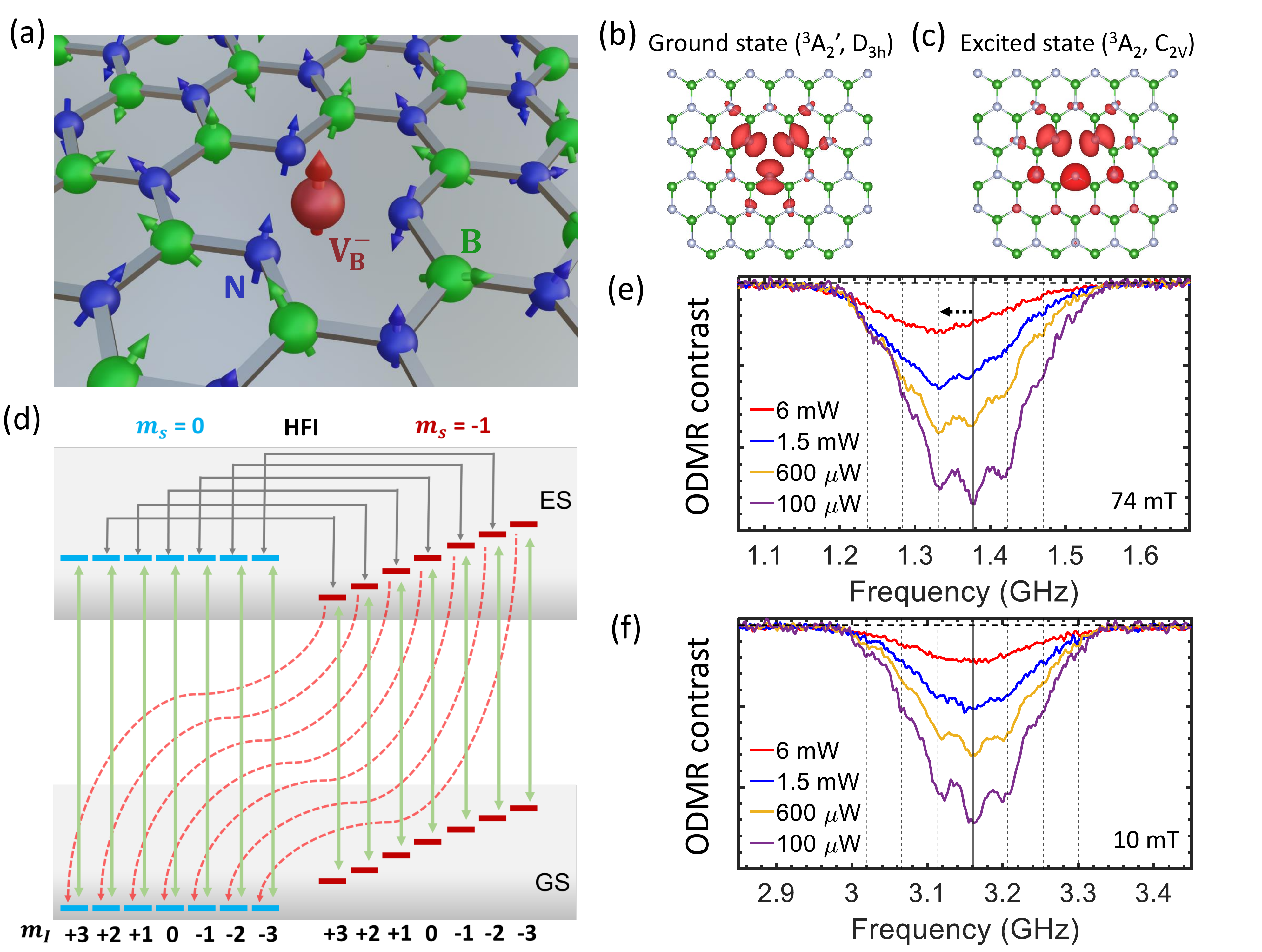}
	\caption{\textbf{Optical polarization of nuclear spins in hBN with V$_B^-$ spin defects.} (a) Illustration of nuclear spins around a V$_B^-$ defect in a 2D hBN lattice. Both nitrogen (blue) and boron (green) atoms have non-zero nuclear spins. The electron spin of the V$_B^-$ defect (red sphere)  couples to the nearest 3 nitrogen nuclear spins via hyperfine interaction. (b) The ground-state electron spin density of a V$_B^-$ defect. (c) The excited-state spin density of a V$_B^-$ defect. (d) Simplified diagram illustrating the dynamics of optical spin polarization at ESLAC. Red dashed lines indicate nonradiative transitions and green solid lines represent optical transitions which conserve nuclear spins. The gray arrows show the transverse HFI that hybridize the electron-nuclear spin states. (e)-(f) ODMR spectra of V$_B^-$ defects at ESLAC (e), and in a weak magnetic field far from ESLAC (f). Dashed lines are guide for eyes, showing the expected positions of hyperfine peaks. The horizontal dashed arrow shows the center shift under laser excitation with different powers. } \label{fig:1}
\end{figure*}

Here we report the first experimental demonstration of optical polarization and coherent control of nuclear spins in a vdW material. We utilize the recently-discovered $V_B^-$ spin defects in hexagonal boron nitride (hBN) \cite{gottscholl2020initialization,gottscholl2021room,abdi2018color,ivady2020ab,reimers2020photoluminescence} to polarize the 3 nearest $^{14}$N nuclear spins around each $V_B^-$ electron spin (Fig.\ref{fig:1}).  hBN has a crystalline structure similar to that of graphene but has a large band gap of 6~eV, making it an ideal vdW material host for optically addressable spin defects \cite{gottscholl2020initialization,chejanovsky2021single,stern2021room}.  So far, the most studied spin defect in hBN is the $V_B^-$  defect \cite{ivady2020ab,reimers2020photoluminescence,abdi2018color}, which can be generated by ion implantation\cite{kianinia2020generation,gao2021high,guo2022generation} and other methods \cite{gottscholl2020initialization,gao2021femtosecond}. $V_B^-$ spin defects have a high optically detected magnetic resonance (ODMR) contrast \cite{gao2021high}, and have been used for quantum sensing \cite{gottscholl2021spin,gao2021high} and quantum imaging of 2D magnetic materials \cite{healey2021quantum,huang2021wide}.
Different from diamond which has sparse nuclear spins \cite{dutt2007quantum}, all atoms in hBN have non-zero nuclear spins. Because they have longer coherence times than those of electron spins, nuclear spins are promising resources for quantum sensing, network, computing and simulation if they can  be polarized and coherently controlled \cite{dutt2007quantum,cai2013large,gangloff2019quantum,ruskuc2022nuclear}.

\begin{figure*}[thbp]
	\centering
	\includegraphics[width=0.6\textwidth]{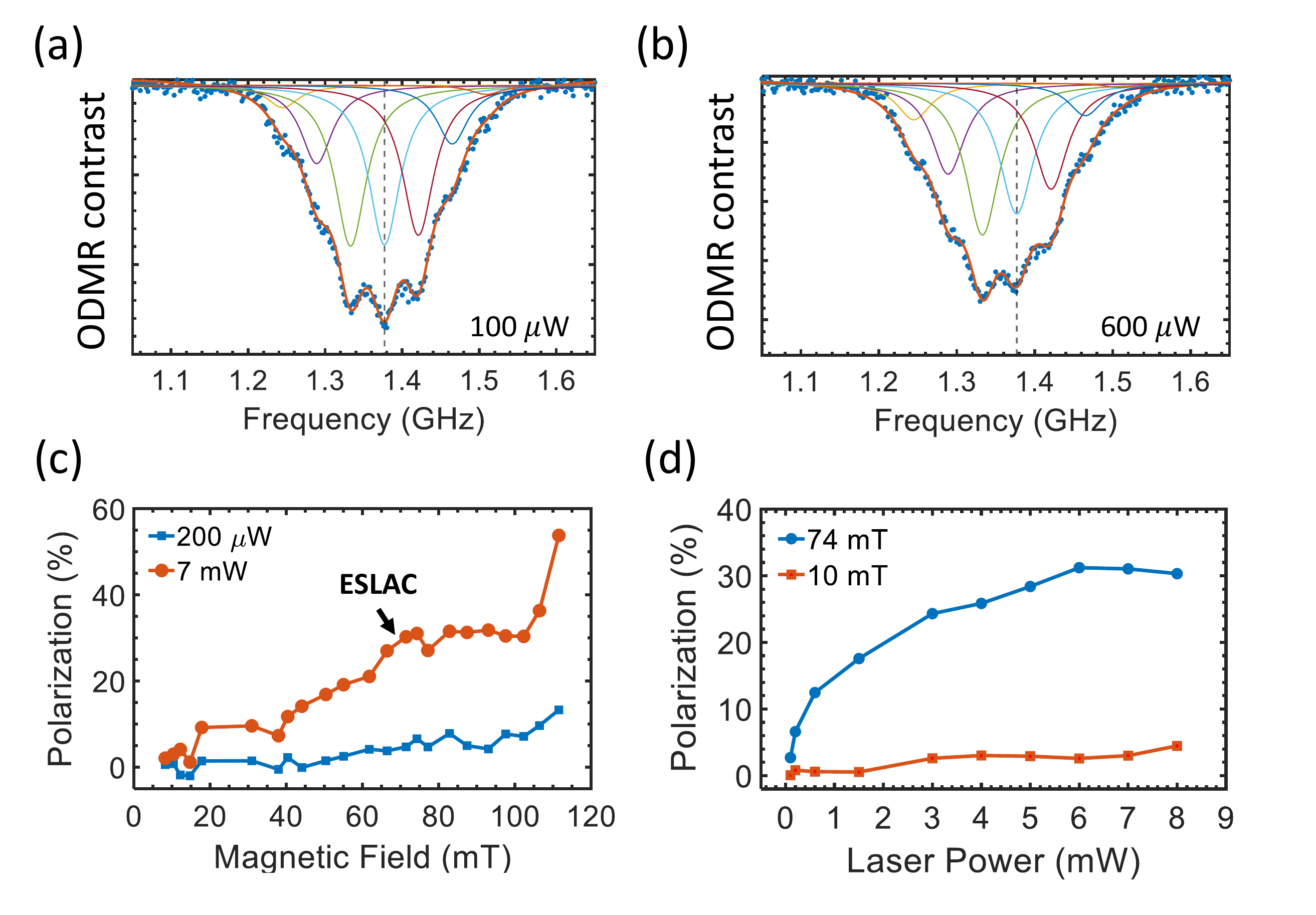}
	\caption{\textbf{Polarization of the three nearest nitrogen nuclear spins.} (a)-(b) ODMR spectrum at ESLAC under (a) 100 $\mu$W laser excitation, and (b) 600 $\mu$W laser excitation. The experimental data is fit using seven Lorentzian curves corresponding to $m_I=+3,+2,+1,0,-1,-2,-3$. The frequency of each hyperfine peak is obtained from the fitting results using the data of 100 $\mu$W, and does not change for higher laser powers. The center dashed vertical line marks the hyperfine peak $m_I=0$. (c) Measured average polarization of the 3 nearest nitrogen nuclear spins as a function of the magnetic field. The nuclear spin polarization increases when the magnetic field increases from 7 mT to 110 mT. A strong laser excitation (7 mW, red curve) produces a  larger polarization than that with a weak laser excitation (200 $\mu$W, blue curve). (d) Nuclear spin polarization as a function of the excitation laser power at ESLAC (74 mT, blue curve) and in a small magnetic field (10~mT, red curve).  } \label{fig:2}
\end{figure*}

In this article, we optically polarize nuclear spins in hBN at room temperature using the hyperfine interaction between nuclear spins and $V_B^-$ electron spins (Fig.\ref{fig:1}). We only consider $^{14}$N nuclei in this work since 99.6$\%$ of natural nitrogen is $^{14}$N, which has spin 1. We use plasmonic enhancement to speed up optical polarization and readout of $V_B^-$ spin defects \cite{gao2021high}. The achieved average polarization of the 3 nearest $^{14}$N nuclear spins around each $V_B^-$ spin defect is about 32$\%$ near the excited state level anti-crossing (ESLAC), and is even larger near the ground state level anti-crossing (GSLAC).
 Thus these nuclear spins are cooled to less than 1 mK with optical pumping when the environment is at room temperature. 
With polarized nuclear spins in hBN, we implement the optically detected nuclear magnetic resonance (ODNMR).  The measured ODNMR spectra of the 3 trigonal nearest nitrogen nuclear spins show strong nuclear-nuclear coupling mediated by electron spins \cite{Bermudez2011PhysRevLett.107.150503} that is $10^5$ times larger than the direct nuclear spin dipolar coupling. We also perform {\it ab initio} calculations \cite{ping2021computational,smart2021intersystem,PhysRevMaterials.1.071001} to support our observations.
Lastly, we drive the nuclear spin transition resonantly and realize coherent control of nitrogen nuclear spin states. Remarkably, the Rabi oscillation of nuclear spins is enhanced by a factor of about 350 near ESLAC due to hyperfine interaction. Utilizing the hyperfine enhancement, we achieve MHz fast coherent control of nuclear spins.

\begin{figure*}[thbp]
	\centering
	\includegraphics[width=1\textwidth]{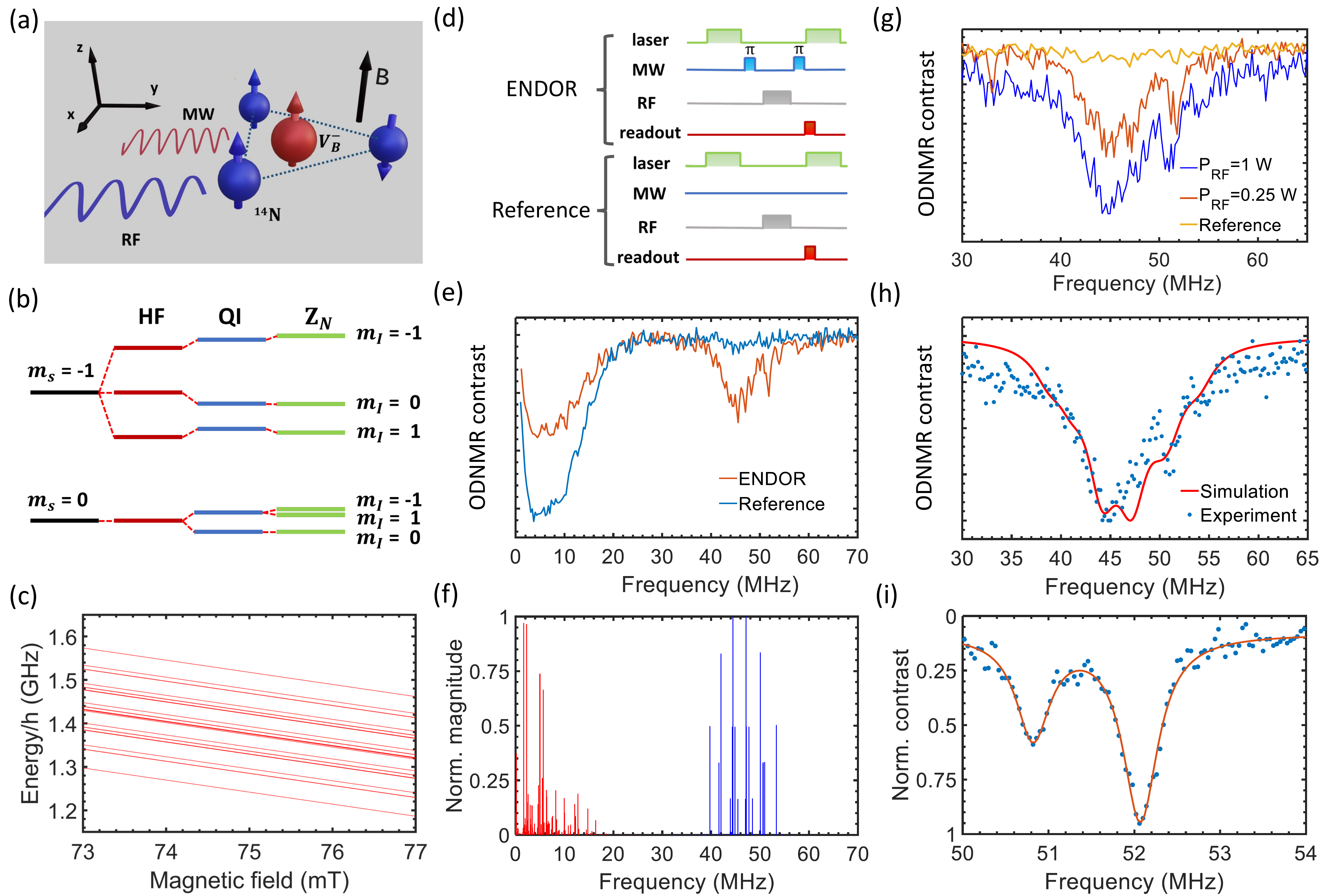}
	\caption{\textbf{ODNMR spectroscopy of the three nearest nitrogen nuclear spins.} (a) Schematic of a V$_B^-$ defect coupled to three nearest nitrogen nuclear spins. An external static magnetic field is applied perpendicular to the hBN flake. A RF pulse generates an in-plane AC magnetic field that drives  nuclear spin transitions. A MW pulse drives electron spin transitions. (b) The energy level diagram  for an electron spin coupled to a nuclear spin in a magnetic field. The interactions include ZFS, electron spin Zeeman effect, HFI, QI and nuclear spin Zeeman effect (Z$_N$). (c) Simulated electron spin energy levels around ESLAC, which has 27 lines for the $m_s=-1$ branch. $h$ in the vertical axis title is the Planck constant. (d) Schematic of the ODNMR Pulse sequences. (e) A large range scan of the ODNMR spectrum. Using the ENDOR sequence, a broad peak around 45 MHz is observed (blue curve), while this peak disappears when there are no MW $\pi$ pulses (red curve). (f) Simulated nulcear spin transitions. (g) A more detailed measurement of the ODNMR spectrum for the $m_s=-1$ branch. (h) Comparison between the experimental result and the numerical simulation, which shows good agreement. (i) Isolating a nulcear spin transition by using a weak RF drive for a longer duration.  } \label{fig:3}
\end{figure*}


As shown in Fig.\ref{fig:1}(a), a $V_B^-$ spin defect is formed by missing a boron atom in the hBN lattice. The $V_B^-$ defect has a spin triplet ground state (GS) with a zero-field splitting (ZFS) of  $D_{GS}=3.45$~GHz \cite{gottscholl2020initialization}, and a spin triplet excited state (ES) with ZFS of $D_{ES}=2.1$~GHz  \cite{mathur2021excited,baber2021excited,yu2021excited,mu2021excited}. The spin-dependent state recombination and photon emission allow  optical initialization and readout of the electron spin state (Supplementary Fig. S1). The  $V_B^-$ electron spin couples to nuclear spins via hyperfine interaction. As the hyperfine interaction with farther nuclear spins are much weaker \cite{ivady2020ab}, we only consider the three nearest $^{14}$N nuclear spins  in this work. The spin system of the $V_B^-$ defects can be described by the same form of the Hamiltonian for both the GS (Fig. \ref{fig:1}(b)) and the ES (Fig. \ref{fig:1}(c), Supplementary Fig. S2-S3), using different parameters. The GS (as well as ES) Hamiltonian in the presence of an external magnetic field $B_0$ includes electron spin-spin interaction (ZFS), electron-nuclear hyperfine interaction (HFI), electron and nuclear Zeeman splitting, and nuclear spin quadrupole interation (QI):
\begin{equation}
\begin{split}
	H = &D[S_z^2-S(S+1)/3]+\sum_{j=1,2,3}\mathbf{S}\mathbf{A}_j\mathbf{I}_j+\gamma_eB_0S_z\\
	&-\sum_{j=1,2,3}\gamma_nB_0I_{zj}+\sum_{j=1,2,3}Q_j(I_{zj}^2-I_j(I_j+1)/3).
\end{split} \label{Eq:1}
\end{equation}
Here $D$ is the ZFS parameter, $\mathbf{S}$ and S$_z$ are the electron spin-1 operators, $\mathbf{I}_j$ and I$_{zj}$ ($j$=1,2,3) are the nuclear spin-1 operators of the three nearest $^{14}$N nuclei, $\mathbf{A}_j$ is the HFI tensor, $\gamma_e = 28$~GHz/T is the electron spin gyromagnetic ratio, $\gamma_n=3.076$~MHz/T is the gyromagnetic ratio of $^{14}$N nuclear spin, and $Q_j$ is the quadrupole coupling constant.  The $z$-axis is perpendicular to the hBN nanosheet.

\begin{figure*}[thbp]
	\centering
	\includegraphics[width=0.6\textwidth]{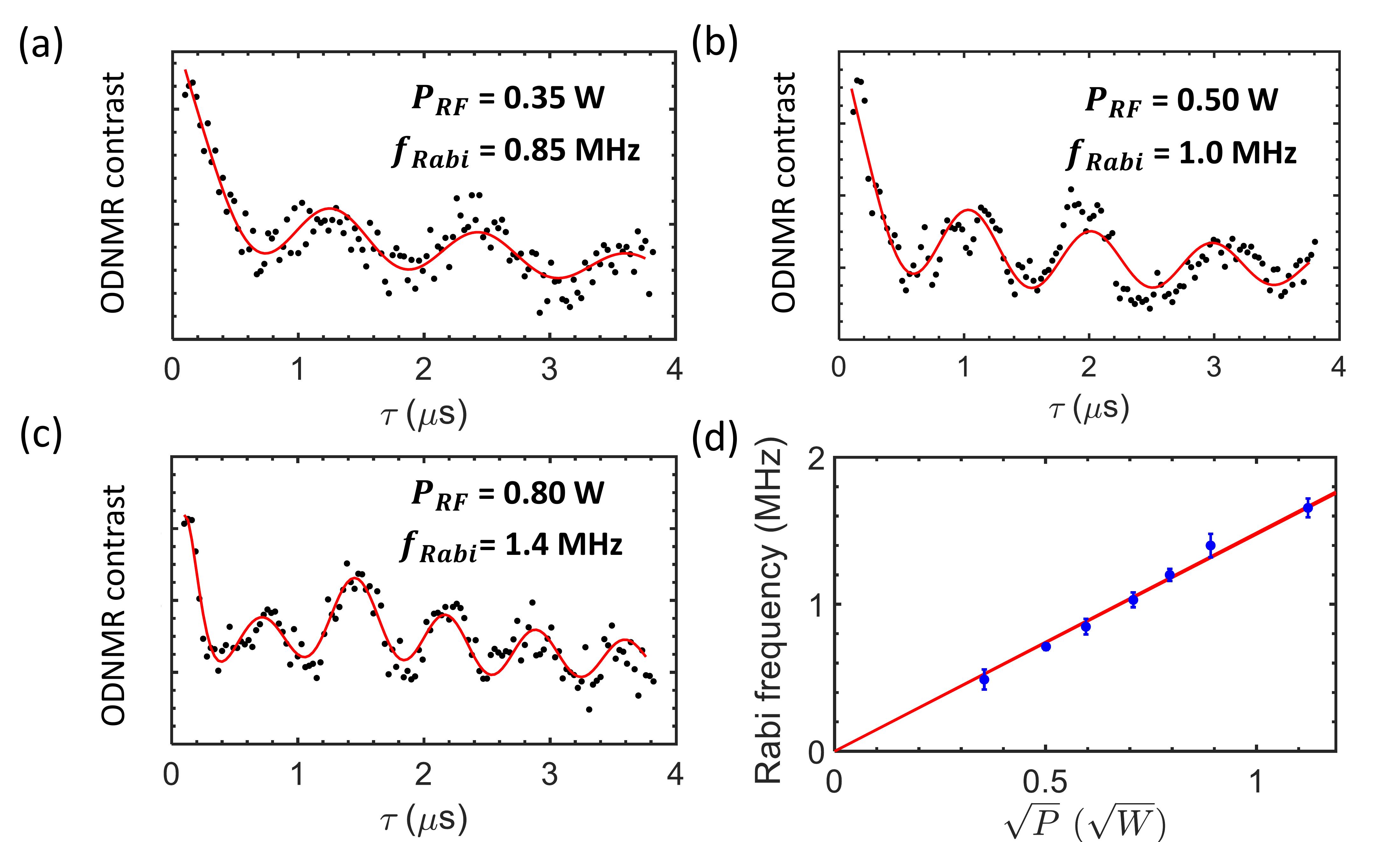}
	\caption{ \textbf{Coherent control of nuclear spins in hBN.} ODNMR contrasts as functions of  RF pulse duration times $\tau$  when the driving RF power is (a) 0.35 W, (b) 0.5 W, and (c) 0.8 W. The frequency of the RF drive is 		
		52.05 MHz.  The magnetic field is 74 mT. The solid lines are fitting results combining a Rabi oscillation and exponential decays. For a strong RF driving (c), the Rabi oscillation contains more than one frequency components while the faster oscillation term dominates. (d) Rabi frequency as a function of the RF driving power. } \label{fig:4}
\end{figure*}

As shown in Fig. \ref{fig:1}(d), in a magnetic field corresponding to the level anti crossing (about 74 mT for the ES and 124 mT for the GS), the sublevels with electron spin $m_s = -1$ approach the  sublevels with electron spin $m_s = 0$. The transverse hyperfine interaction then hybridizes the state $\ket{m_s=0,m_I}$ and the state $\ket{m_s=-1,m_I+1}$ (here we denote $m_I$ as the total $z$ component of the three nearest $^{14}$N nuclei). Continuous optical pumping keeps initializing the electron spin state into the $m_s=0$ state while conserving the nuclear spin $m_I$. Meanwhile, by means of the state mixing, the total spin state will have a chance to evolve into $\ket{m_s=-1,m_I+1}$ in each optical cycle \cite{PhysRevLett.102.057403}. In the ideal case, the  spin system  will be polarized into an unmixed $\ket{m_s=0,m_I=3}$ state eventually.

In the experiment, we use a 532 nm laser to polarize nuclear spins at room temperature and use  ODMR to measure the nuclear spin distribution.  First, we use a low power laser (100~$\mu$W) for spin initialization and readout (Fig. \ref{fig:1} (e), \ref{fig:2}(a)), which avoids power broadening and has a mild effect on nuclear spin polarization. 
As a result, the hyperfine structure is clearly resolved and the ODMR spectrum is almost symmetric around the center peak ($m_I=0$), indicating a low nuclear spin polarization.
  When we increase the laser power, significant distortion and shift of the ODMR spectrum are observed at 74 mT (Fig. \ref{fig:1} (e), \ref{fig:2}(b)), indicating large polarization of nuclear spins. In contrast, in a 10 mT magnetic field,  the center remains nearly at the same position under different laser power excitation (Fig. \ref{fig:1} (f)).

The  average polarization of the 3 nearest $^{14}$N nuclear spins is measured as
$P_{exp}=\sum_{m_I} m_I \rho_{m_I}/(3\sum_{m_I} \rho_{m_I})$,
where the summation is performed over the seven hyperfine peaks in the ODMR spectrum. $\rho_{m_I}$ denotes the fitted relative population of $m_I$ states (Fig. \ref{fig:2} (a)) (see Methods for more details). 
Fig. \ref{fig:2}(c) shows the nuclear spin polarization $P_{exp}$ as a function of magnetic fields. In a weak magnetic field that is far from ESLAC, $P_{exp}$ is small. Around ESLAC, $P_{exp}$ increases when the the laser power increases, and reaches 32$\%$ under a high laser power excitation (6 mW) (Fig. \ref{fig:2}(d)). This polarization would require a magnetic field of about $10^6$~T if they were polarized by thermal distribution at room temperature.  The nuclear spin polarization increases further and exceeds 50$\%$ near the GSLAC. Numerical simulation of the optical polarization process can be found in Supplementary Information (Fig. S4-S5).

With polarized nuclear spins,  we conduct ODNMR experiments to gain more insight on the coupled electron-nuclear spin system in hBN (Fig. \ref{fig:3}). We first implement the electron nuclear double reasonce (ENDOR) technique to obtain the ODNMR spectroscopy of the three nearest $^{14}$N nuclei (Fig. \ref{fig:3}(d)). After initializing the system into the $\ket{m_s=0,m_I}$ state using a 7 mW  laser pulse, a selective MW $\pi$ pulse is applied on the electron spin to transfer the population to the $\ket{m_s=-1,m_I}$ state. Then we use a  RF pulse to  drive nuclear spin transitions to change the nuclear spin state from $m_I$ to $m_I'$.   Finally, optical readout is performed after applying another MW $\pi$ pulse that transfers the population back from $\ket{m_s=-1,m_I'}$ to  $\ket{m_s=0,m_I'}$. 
The ODNMR spectrum is presented in Fig. \ref{fig:3}(e). We observe a broad peak around 45 MHz due to nuclear spin transitions among the $\ket{m_s=-1,m_I}$ states. The center of the peak is close to the hyperfine interaction constant $A_{zz}=47$ MHz. Meanwhile, there is another broad peak around 5 MHz due to nuclear spin transitions among $\ket{m_s=0,m_I}$ states. To confirm our observation, we perform another ODNMR measurement  without the MW pulse as the ``Reference'' (Fig. \ref{fig:3} (d)-(e)). Under this condition, the electron spin  stays in the $m_s=0$ state after the laser initialization. So there is no hyperfine interaction. As a result, the signal around 45 MHz in the reference ODNMR spectrum is negligible. Meanwhile, the magnitude of the peak around 5 MHz increases because more electron spins stay at $m_s=0$ state.

We support our experimental findings by modeling the system using the full Hamiltonian (Eq. \ref{Eq:1}) that consists of a $V_B^-$ electron spin and the three nearest $^{14}$N nuclear spins. Due to the hyperfine interaction between 3 $^{14}$N nuclear spins and the electron spin, there are  27  energy sublevels for the $m_s=-1$ branch as shown in Fig. \ref{fig:3}(c). 
 As presented in Fig. \ref{fig:3}(f), these sublevels result in many allowed nuclear spin transitions over a broad frequency range. This causes broadening  in our measured ODNMR spectrum (Fig. \ref{fig:3}(g)). Our simulated NMR spectrum agrees well with experimental results as shown in Fig. \ref{fig:3}(h) (see more details in the Supplementary Information, Fig. S6-S8).  With a weaker but longer RF pulse, we can resolve a narrow  peak near 52 MHz (Fig. \ref{fig:3}(i)).
 
 The broad distribution of allowed transitions in ODNMR is due to the strong nuclear-nuclear coupling mediated by the electron spin \cite{Bermudez2011PhysRevLett.107.150503}.  
 These three $^{14}$N nuclear spins are strongly coupled to the same electron spin via  hyperfine interaction. When the system is far away from the GSLAC ($|D_{GS}-\gamma_e B|\gg A_{xx}, A_{yy}, A_{zz} $), the effective nuclear-nuclear spin coupling constant is  $C_{NN}=A^2_{tran}/|D_{GS}-\gamma_e B|$ for the $m_s=-1$ branch, where $A_{tran}=(A_{xx}+A_{yy})/2=68$~MHz is the transverse hyperfine interaction constant (Supplementary Information).   
 At 74~mT, $C_{NN}=3.4$~MHz which is $10^5$ times larger than the direct nuclear spin dipolar coupling constant $d_{NN}=\mu_0 (\gamma_n)^2 \hbar/(2 r^3_{NN})=34$~Hz. Here $\mu_0$ is the vacuum permeability, $\hbar=h/(2\pi)$, and $r_{NN}$ is the separation between two nearest nitrogen nuclei. $C_{NN}$ is large near GSLAC, but decreases when the magnetic field is very large. $C_{NN}=50$~kHz at 3.3~T. Thus 
the profiles of our measured ODNMR spectra at 74~mT are very different from former results obtained at 3.3~T using a commercial pulsed ESR spectrometer (See detailed numerical analysis in the Supplementary Information) \cite{murzakhanov2021electron}.  Because of small $C_{NN}$ in a large magnetic field, the former work with V$^-_B$ spin defects at 3.3 T did not observe the nuclear-nuclear spin interaction \cite{murzakhanov2021electron}.  Our observation of strong MHz nuclear-nuclear coupling will be important for mutliqubit quantum gates.

 We now perform coherent control of nuclear spins in hBN. For an isolated $^{14}$N nucleus, its Rabi frequency will be $\gamma_e/\gamma_n=9110$ times smaller than that of an electron spin (Supplementary Fig. S9), making it challenging to perform coherent control. However, the  $V_B^-$ electron spin can increase the Rabi frequency of the nearby nuclear spins by  hyperfine enhancement.  Because of the large  $\gamma_e/\gamma_n$ ratio, even a slight coupling can lead to a large enhancement in the Rabi frequency \cite{chen2015measurement,Sangtawesin_2016}. We will use the $m_s=-1$ electron state because it gives a larger hyperfine enhancement (see Supplementary Information for more details).
We observe Rabi oscillations by performing ODNMR experiments using a 52 MHz RF drive with a varying pulse length. Fig. \ref{fig:4}(a),(b),(c) illustrate the Rabi oscillations of the nuclear spin state under different RF driving powers. By fitting the Rabi oscillation, we estimate the inhomogeneous coherence time $T^*_{2n}$ of nuclear spins to be about  3.5 $\mu$s at room temperature. This $T^*_{2n}$ is much longer than the inhomogeneous coherence time $T^*_{2e}$  of $V_B^-$ electron spins, which is about 100 ns (Supplementary Information). The Rabi frequency shows a good linear dependence on the square root of the RF power (Fig. \ref{fig:4}(d)). By comparing the Rabi frequencies of the electron spin and the nuclear spin, we find the Rabi frequency of the  $^{14}$N nuclear spin is enhanced by a factor of 350 at ESLAC, which enables fast coherent control  (Supplementary Fig. S10). A simplified theoretical model predicts the hyperfine enhancement of nuclear spins coupled to  $m_s=-1$ electron state to be about 420 at ESLAC (Supplementary Information). Thus the experimental result shows good agreement with theoretical prediction.

In conclusion, we have optically polarized nuclear spins in a van der Waals material with intrinsic electron spin defects. By making use of ESLAC and GSLAC of V$_B^-$ spin defects in hBN, we are able to polarize the three nearest $^{14}$N nuclear spins at room temperature over a broad range of magnetic fields. Our ODNMR measurements show the first NMR spectrum using the intrinsic spin defects of hBN. This further reveals the strong nuclear-nuclear spin coupling mediated by the electron spin, which will enable multiqubit operations. We also demonstrate MHz fast coherent control of the nuclear spins with hyperfine enhancement. The polarized nuclear spins in vdW materials will have potential applications in quantum sensing, network, computing and simulation \cite{liu20192d,kane1998silicon,cai2013large,dutt2007quantum,gangloff2019quantum,ruskuc2022nuclear}. 
Nitrogen nuclear spins in the triangle lattice in hBN will be suitable for large-scale quantum simulation of different magnetic states \cite{cai2013large}, including spin liquids \cite{banerjee2016proximate,chen2020strong}. Their coherence time can be improved further by reducing the temperature and engineering isotope compositions \cite{lee2022first,haykal2021decoherence}.




%

\end{document}